\documentclass[journal,comsoc]{IEEEtran}
\ifCLASSINFOpdf
\else
\fi
\usepackage{amsfonts,amsmath,amssymb}
\usepackage{mathtools}
\usepackage{mdwmath}
\usepackage{cuted}
\usepackage{mdwtab}
\usepackage{bbold}
\usepackage{amsthm}           % for designing theorems
\usepackage{bm}               % provides bold math \bm{} command (vectors)
\usepackage{units}            % for nicely typset units \unit[10]{kHz}
\usepackage{graphicx}
\usepackage[linesnumbered,ruled,vlined]{algorithm2e}

\usepackage{url}
\usepackage{paralist}
\newcommand{\Fig}[1]{Fig.~\ref{fig:#1}}

\newcommand{\Eq}[1]{Eq.~(\ref{eq:#1})}

\newcommand{\Lc}{\mathcal{L}}

\newcommand{\Rs}{\mathbb{R}^2}

\newcommand{\Ed}{\mathbb{E}}

\newcommand{\Pd}{\mathbb{P}}
\newcommand{\sinr}{\mathrm{SINR}}

\newcommand{\dr}{\mathrm{d}}
\newcommand{\Rcone}{u(\omega,\gamma)}

\newcommand{\Upper}{u}
\newcommand{\Lower}{l}


\DeclarePairedDelimiter\floor{\lfloor}{\rfloor}

\begin{document}
\title{Coverage Analysis for Low-Altitude UAV Networks in Urban Environments}

\author{Boris Galkin,
        Jacek~Kibi\l{}da,
        and~Luiz~A. DaSilva% <-this % stops a space
\thanks{This material is based upon works supported by the Science Foundation
Ireland under Grants No. 10/IN.1/I3007 and 14/US/I3110. B. Galkin, J. Kibi\l{}da, and L. DaSilva are with CONNECT, Trinity College Dublin, Ireland, email:  \{galkinb,kibildj,dasilval\}@tcd.ie.}% <-this % stops a space
}

\maketitle

\begin{abstract}
Wireless access points on unmanned aerial vehicles (UAVs) are being considered for mobile service provisioning in commercial networks. To be able to efficiently use these devices in cellular networks it is necessary to first have a qualitative and quantitative understanding of how their design parameters reflect on the service quality experienced by the end user. In this paper we set up a scenario where a network of UAVs operating at a certain height above ground provide wireless service within coverage areas shaped by their directional antennas. We provide an analytical expression for the coverage probability experienced by a typical user as a function of the UAV parameters. 
\end{abstract}


% \fi
\IEEEpeerreviewmaketitle

\section{Introduction}
To meet growing data demands and new technologies emerging in the telecommunications sector, operators and over-the-top service providers are considering the use of unmanned aerial vehicles (UAVs) for delivering wireless service. These wireless-provisioning UAV platforms vary greatly in size and operating range, with high-altitude UAVs operating across hundreds of kilometers at altitudes previously reserved for manned aircraft on one end and miniature quadcopter-style UAVs with ranges of a few hundred meters on the other. UAVs in the latter category are in particular drawing the attention of the internet of things (IoT) community: the authors of \cite{Motlagh_2016} suggest that the majority of UAVs in IoT applications will be miniature devices that operate at heights below 300 meters. The reason for this is that miniature, low altitude UAVs offer lower cost, more flexible deployment and they make use of airspace which is far less utilised by manned aircraft and is therefore subject to more relaxed regulations \cite{Atkins_2010}. 
%Following this developing interest in low-altitude UAV networks there is a growing necessity for flexible and robust modelling tools which can accurately describe the unique benefits and challenges of operating in populated areas at such low heights. Given the suggested height ceiling the UAVs in the network may be operating far above a built-up urban area or below building heights in so-called urban canyons, the effect of this transition on the network performance and its implications for UAV network optimisation must be modelled and understood.

The UAVs in the network may be operating far above a built-up urban area or below building heights in so-called urban canyons. In this paper we employ stochastic geometry to model the effect of this transition on coverage probability and its implications for UAV network optimisation. Our model of the environment takes into account parameters such as building density and UAV antenna beamwidth, and can represent different wireless fast-fading behaviours through generalised Nakagami-$m$ fading. As we simulate actual building distributions, our model affords us higher precision when analysing UAV deployments with cell sizes comparable to terrestrial picocells, with UAVs hovering at heights of around 100 meters and exhibiting coverage range of the order of a couple of hundred meters. In addition to this improved level of granularity, our model retains generality, as its applicability is not restricted to particular environments or UAV placement strategies. We demonstrate with our model how the line-of-sight (LOS) blocking effect of buildings can have either a beneficial or a detrimental effect on coverage probability, depending on the density of the UAV network and the coverage probability threshold under consideration. We also demonstrate that, for a given beamwidth of the UAV antenna, there exists an optimum height which maximises the coverage probability for a given UAV density and coverage probability threshold. To our knowledge, we are the first to combine high-detail environment models with stochastic geometry to analyse coverage probability in low-altitude UAV networks.

\subsection{Related Work}
The wireless community has published several works on the modelling of wireless links in UAV networks operating at altitudes in the order of kilometers. In \cite{AlHourani_20142} the authors use an ITU model to describe an urban environment and then apply raytracing simulations to describe the pathloss from LOS and non-line-of-sight (NLOS) components. The same authors in \cite{AlHourani_2014} model the probability of a UAV having an LOS channel to a user as a sigmoid function of the vertical angle between the UAV and user; they then use this model to describe the coverage radius of the UAV as a function of pathloss and demonstrate how the UAV height can be optimised through maximising the coverage radius in an interference-free environment. In \cite{Mozaffari_2015, Hayajneh_2016} the authors use this sigmoid function LOS model to optimise UAV height for different performance metrics, and in \cite{Mozaffari_20162, Fotouhi_2016} the authors apply multi-objective optimisation to UAV networks, using the sigmoid LOS model to characterise the received signal strength. Note that in the above works the UAV locations are assumed to either be known $a$ $priori$ or are found as part of an optimisation problem.

 Stochastic geometry is an alternative method for modelling the spatial relationships in a UAV network. Without prior knowledge of the UAV locations, it is possible to describe the UAVs as being distributed in space randomly, as a function of geometric parameters such as UAV density. This approach is followed in \cite{Ravi_2016} and \cite{Chetlur_2017}, in which the authors derive the coverage probability for a stochastic UAV network under guaranteed LOS conditions for a fading free and Nakagami-$m$ fading channel. The authors describe a fixed number of UAVs operating within a fixed area at a certain height above ground and demonstrate how an increase in height results in a decrease in the coverage probability. Additionally in \cite{Chetlur_2017} they demonstrate how larger values of fading parameter $m$ reduce the variance of the random signal-to-interference ratio (SIR) experienced by the user. Stochastic geometry is also applied by the authors of \cite{Zhang_2017} to optimise UAV density in a radio spectrum sharing scenario under guaranteed LOS conditions.
 
 As the works above consider high-altitude UAV networks they simplify the UAV-user link, either by applying a basic vertical angle-based LOS model or by assuming guaranteed LOS channels outright. These simplifications may be reasonable for high-altitude UAV networks; however, in a scenario where the UAVs may operate below building heights we expect the buildings to have a much more tangible effect on the network. As such, the models applied in the state-of-the-art may be too simplistic for the low-altitude UAV network scenario under consideration.

\begin{figure}[t!]
\centering
	\includegraphics[width=.40\textwidth]{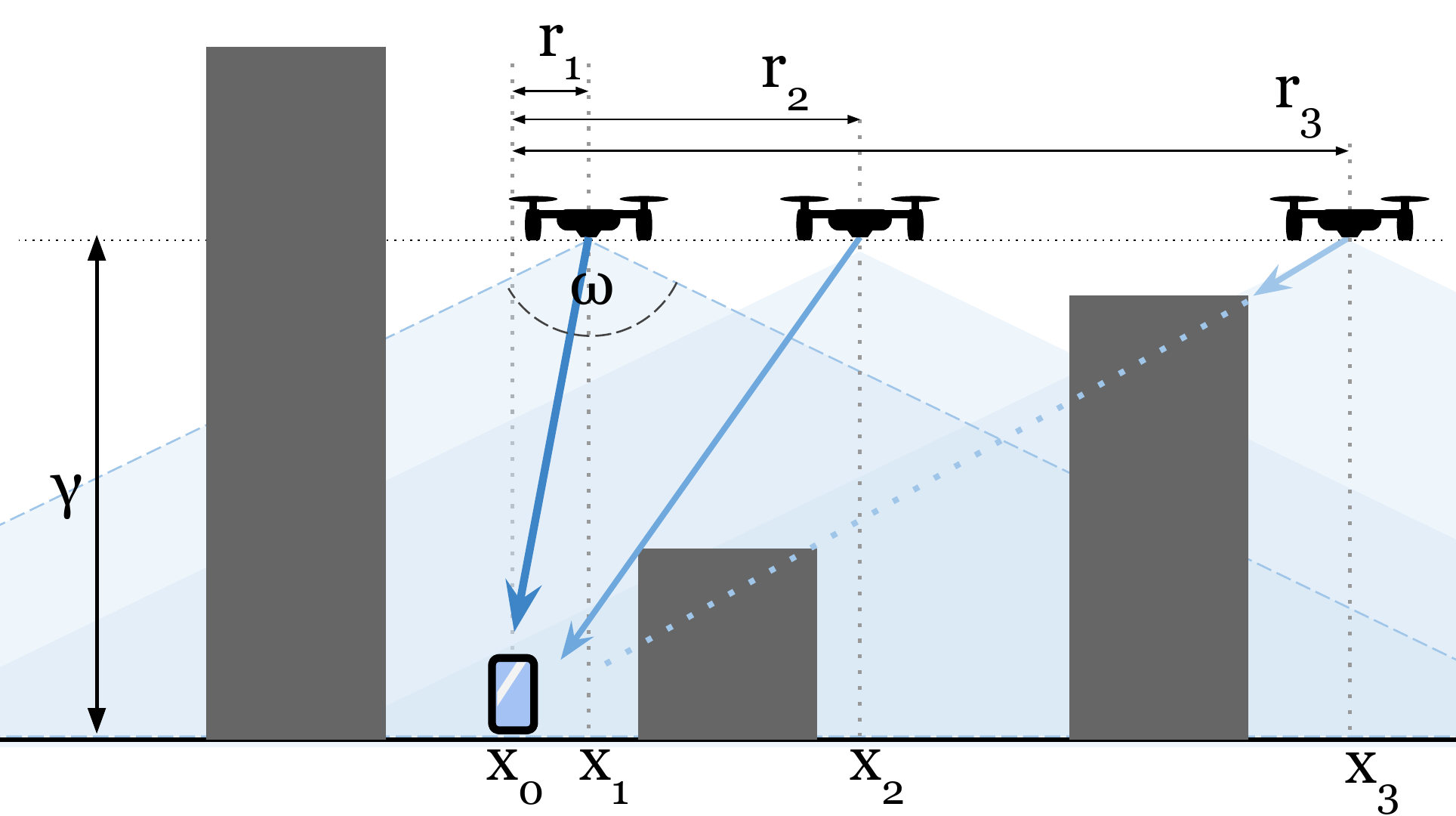}
	\caption{
	Side view showing UAVs in an urban environment at a height $\gamma$, with 2D coordinates $x_1,x_2$, $x_3$ and antenna beamwidth $\omega$. The user is serviced by the UAV with the strongest signal, while the remaining UAVs generate LOS and NLOS interference.
	\vspace{-5mm}
	}
	\label{fig:drone_network}
\end{figure}

\section{System Model}

\Fig{drone_network} illustrates our system model. We consider a network of UAVs at a height $\gamma$ above ground and a reference user. We position the reference user at $x_0\in\Rs$ and model the network of UAVs as a point process $\Phi = \{x_1 , x_2 , ...\} \subset \Rs$, where elements $x_i\in \Rs$ represent the projections of the UAV locations onto the $\Rs$ plane. We denote the horizontal distance between the user and a UAV $i$ as $r_i = ||x_i-x_0||$. UAVs have identical capabilities: transmit power $p$ and a directional antenna with beamwidth $\omega$. The main beam illuminates the area directly beneath the UAV. This coverage cone has a radius of $\Rcone=\tan(\omega/2)\gamma$. We assume a uniform and rotationally symmetric beam pattern; using the approximations (2-26) and (2-49) in \cite{Balanis_2005} and assuming perfect radiation efficiency the antenna gain can be expressed as $\eta(\omega) = 16\pi/(\omega^2)$ inside the coverage cone and $\eta(\omega)=0$ outside of the coverage cone. It follows that a reference user will receive a signal from a UAV $i$ and only if $r_i \leq \Rcone$. 

We define a bounded circular area $\mathcal{W}\subset \Rs$ of radius $\Rcone$ centered at the reference user. We represent the UAV network with a homogeneous Poisson point process (PPP) with intensity $\lambda$, so the UAVs in the area $\mathcal{W}$ form a point process $\Phi_{\mathcal{W}} = \{x_i\in\Phi : r_i \leq \Rcone\}$ that is also a PPP with the same intensity $\lambda$. Given that a PPP is translation invariant with respect to the origin, we can set the 2D coordinates of our reference user to $x_0=(0,0)$. 

Both the serving UAV and the interfering UAVs will be affected by buildings in the environment, which form obstacles and break LOS links. We adopt the model in \cite{ITUR_2012}, which defines an urban environment as a collection of buildings arranged in a square grid. There are $\beta$ buildings per square kilometer, the fraction of area occupied by buildings to the total area is $\delta$, and each building has a height which is a Rayleigh-distributed random variable with scale parameter $\kappa$. The probability of a UAV $i$ having LOS to the reference user is given in \cite{ITUR_2012} as 

\begin{equation}
\Pd_{LOS}(r_i) = \prod\limits_{n=0}^{\max(0,d_i-1)}\left(1-\exp\left(-\frac{\left(\gamma - \frac{(n+1/2)\gamma}{d_i}\right)^2}{2\kappa^2}\right)\right)
\label{eq:LOS}
\end{equation}
%Let us denote the distance to UAV $i$ as $r_i = ||x_i||$. The new set $\Phi_{\mathcal{W}}' = \{r_i \in \Rb\}$ is an inhomogeneous Poisson point pattern with intensity $\lambda'(r) = 2\pi\lambda r$. 

\noindent
where $d_i = \floor*{r_i\sqrt{\beta\delta}}$. 

Let $S_{i}$ be the power received from the $i$th UAV by the reference user; for a given value of $r_i$ this is defined as $S_{i} = p\eta(\omega) H_{t_i} (r_i^2+\gamma^2)^{-\alpha_{t_i}/2}$ where $H_{t_i}$ is the random multipath fading component, and $\alpha_{t_i}$ is the pathloss exponent, where $t_i \in \{\text{L},\text{N}\}$ is an indicator variable which denotes whether the $i$th UAV has LOS or NLOS to the user. The signal-to-interference-and-noise ratio (SINR) for the reference user can be described as $\sinr = S_{1}/(I_L + I_N+\sigma^2)$ where $S_1$ denotes the serving UAV signal, $I_L$ and $I_N$ denote the aggregate LOS and NLOS interference and $\sigma^2$ denotes the noise power.

The reference user is said to be covered by the UAV network if two conditions are met. First, there exists at least one UAV within $\mathcal{W}$ such that the user is inside of a UAV coverage cone from at least one of the UAVs. The probability of this occurring is given as 

\begin{equation}
\Pd(|\Phi_{\mathcal{W}}|>0) = 1- \exp(-\pi\lambda\Rcone^2),
\label{eq:assProb}
\end{equation}

\noindent
where $|.|$ denotes set cardinality. Second, the SINR experienced by the user is above some threshold $\theta$. The reference user's coverage probability is derived in the following section.

\section{Analytical Results}

As we are interested in the distances of the UAVs to the reference user rather than their 2D coordinates in $\mathcal{W}$ we can apply the mapping theorem \cite{Haenggi_2013}[Theorem 2.34] to convert the 2D PPP $\Phi_{\mathcal{W}}$ into a 1D PPP $\Phi^\prime_{\mathcal{W}} \subset [0,\Rcone]$. $\Phi^\prime_{\mathcal{W}}$ is an inhomogeneous PPP with intensity function $\lambda'(r) = 2\pi\lambda r$ where the coordinates of the UAVs correspond to their horizontal distances to the user. Note that we drop the index $i$ as the horizontal distances of the UAVs in $\Phi^\prime_{\mathcal{W}}$ have the same distribution irrespective of their index values. We partition $\Phi^\prime_{\mathcal{W}}$ into two PPPs which contain the LOS and NLOS UAVs, denoted as $\Phi^\prime_{\mathcal{W}L} \subset [0,\Rcone]$ and $\Phi^\prime_{\mathcal{W}N} \subset [0,\Rcone]$, respectively, with intensity functions $\lambda'_{L}(r) = \Pd_{LOS}(r) 2\pi\lambda r$ and $\lambda'_{N}(r) = (1-\Pd_{LOS}(r)) 2\pi\lambda r$. In effect, the LOS probability function acts as a thinning function \cite{Haenggi_2013} which removes (thins) UAVs from the PPP $\Phi^\prime_{\mathcal{W}}$ with probability $(1-\Pd_{LOS}(r))$ to form $\Phi^\prime_{\mathcal{W}L}$ from the remaining UAVs and $\Phi^\prime_{\mathcal{W}N}$ from those that are thinned.

\subsection{Distribution of the Distance to the Serving UAV}
The received signal power is affected by the distance between the UAV and the reference user, the multipath fading $H_{t_i}$ and the pathloss $\alpha_{t_i}$. Note that $\alpha_N > \alpha_L$ to represent the attentuation that happens when a wireless signal encounters obstacles. As a result of this difference in pathloss, UAVs which are physically closer to the reference user but that are blocked by buildings will have lower received signal strength than UAVs which are further away but have LOS to the user. This introduces a complication to determining the serving UAV for the reference user. In the literature, when UAVs are assumed to all have identically performing wireless channels to the user the UAV that is closest to the user will also be the UAV with the strongest received signal power; therefore, the user will always be serviced by the closest UAV, and the UAVs beyond the serving UAV distance act as interferers. To account for the LOS blocking effects we adopt a different approach to determining which UAV a user associates with. The user will be serviced by the UAV that provides the strongest received signal power. If the multipath fading effect $H_{t_i}$ is averaged out the strongest signal power will come from \textit{either} the closest LOS UAV to the user \textit{or} the closest NLOS UAV. We denote the horizontal distance to the serving UAV as the random variable $R_1$ and refer to it as the serving UAV distance. If a user is serviced by an LOS UAV a horizontal distance $r_1$ away then this means that there are no LOS UAVs with horizontal distance smaller than $r_1$ to the user and that there are no NLOS UAVs with horizontal distance smaller than some lower distance bound $b_N$, where
 
\begin{align}
&p\eta(\omega)(r_1^2+\gamma^2)^{-\alpha_{L}/2} = p\eta(\omega)(b_N^2+\gamma^2)^{-\alpha_{N}/2}, \nonumber \\
 & b_N = \sqrt{\max(0,(r_1^2+\gamma^2)^{\alpha_L/\alpha_N} - \gamma^2)}.
\end{align}
  \noindent
  Note that if $b_N=0$ this means that the LOS serving UAV is close enough to the user that no NLOS UAV will be able to provide a stronger signal no matter how close to the user. 
  
  We can now derive the expression for the probability distribution of the serving UAV distance $R_1$ when the serving UAV is LOS by combining the probability that the closest LOS UAV in the PPP $\Phi^\prime_{\mathcal{W}L}$ is at $r_1$ (as given in \cite{Haenggi_2013}) with the probability that no NLOS UAV exists within a distance $b_N=\sqrt{\max(0,(r_1^2+\gamma^2)^{\alpha_L/\alpha_N} - \gamma^2)}$, giving $f_{R_{1},t_1}(r_1,t_1 = \text{L})$ as
  
  \begin{align}
  %& f_{R_{1},t_1}(r_1,t_1 = \text{L}) = \nonumber \\
  &\Pd_{LOS}(r_1) 2\pi\lambda r_1\exp\left(-2 \pi \lambda \int\limits_{0}^{r_1}\Pd_{LOS}(r)r\dr r\right) \nonumber \\ &\cdot\exp\left(-2\pi\lambda\int\limits_{0}^{b_N}\big(1-\Pd_{LOS}(r)\big)r\dr r\right).
  \label{eq:pdfLOS}
  \end{align}

\noindent
The probability distribution for the distance to the serving UAV when it has NLOS to the user can be obtained following the same logic as above, to give $f_{R_{1},t_1}(r_1, t_1 = \text{N})$ as

\begin{align}
%&f_{R_{1}}(r_1 | t_1 = \text{N}) = \nonumber \\
&\big(1-\Pd_{LOS}(r_1)\big) 2\pi\lambda r_1\exp\left(-2 \pi \lambda \int\limits_{0}^{r_1}\big(1-\Pd_{LOS}(r)\big)r\dr r\right) \nonumber \\ &\cdot \exp\left(-2\pi\lambda\int\limits_{0}^{b_L}\Pd_{LOS}(r)r\dr r\right),
\label{eq:pdfNLOS}
\end{align}

\noindent
where $b_L = \min(\Rcone,\sqrt{(r_1^2+\gamma^2)^{\alpha_N/\alpha_L} - \gamma^2})$ is the lower bound on the horizontal distances of the LOS UAVs. Note that if $b_L = \Rcone$ this denotes that all LOS UAVs must be outside of the window $\mathcal{W}$ for the user to receive the strongest signal from an NLOS UAV at $r_1$. 

 Having obtained the probability distributions of the serving UAV distance we can calculate the probability that a user will have an LOS channel to the UAV that is serving it as  
 
 \begin{equation}
 \Pd(t_1 = \text{L}||\Phi_{\mathcal{W}}|>0) = \frac{\int\limits_{0}^{\Rcone}f_{R_{1},t_1}(r_1,t_1 = \text{L})\dr r_1}{\Pd(|\Phi_{\mathcal{W}}|>0)}.
 \label{losServ}
 \end{equation}

\subsection{Aggregate LOS \& NLOS Interference}

Having defined the lower bounds on the horizontal distances at which the LOS and NLOS interfering UAVs may be found, we can now characterise the expressions for the aggregate LOS and NLOS interference. The LOS and NLOS interferers will belong to the sets $\Phi^\prime_{\mathcal{W}L} \setminus [0,b_L(t_1)]$ and $\Phi^\prime_{\mathcal{W}N} \setminus [0,b_N(t_1)]$, where $b_L(t_1)$ and $b_N(t_1)$ denote the lower bounds on LOS and NLOS interferer distances, as functions of the serving UAV type. If the serving UAV is LOS then $b_L(\text{L}) = r_1$ and $b_N(\text{L}) = \sqrt{\max(0,(r_1^2+\gamma^2)^{\alpha_L/\alpha_N} - \gamma^2)}$, and if the serving UAV is NLOS then $b_L(\text{N}) =  \min(\Rcone,\sqrt{(r_1^2+\gamma^2)^{\alpha_N/\alpha_L} - \gamma^2})$ and $b_N(\text{N}) = r_1$. The aggregate LOS and NLOS interference is then described as $I_L=\sum_{r\in\Phi^\prime_{\mathcal{W}L}\setminus [0,b_L(t_1)]}p\eta(\omega) H_L (r^2+\gamma^2)^{-\alpha_L/2}$ and $I_N=\sum_{r\in\Phi^\prime_{\mathcal{W}N}\setminus [0,b_N(t_1)]}p\eta(\omega) H_N (r^2+\gamma^2)^{-\alpha_N/2}$. 
 % and $\sigma$ is the noise power.

%We assume that all of the interferers have LOS on the UAV.%UAVs beyond the distance $r_{max}$ will not generate interference to the user, UAVs closer than $r_{max}$ will generate interference and will be either LOS or NLOS to the user. We partition the UAV point pattern consisting of the $D-1$ non-servicing UAVs $\Phi\backslash s$ into three disjoint point patterns $\Phi_{LOS}, \Phi_{NLOS},\Phi_{non}$ representing LOS interfering UAVs, NLOS interfering UAVs and non-interfering UAVs, respectively. We define $\textbf{L} = |\Phi_{LOS}|$ as the number of LOS interfering UAVs, $\textbf{N}= |\Phi_{NLOS}|$ the number of NLOS interferers and $\textbf{I} =\textbf{L}+\textbf{N}$ as the total interferer count. Due to the random distribution of UAVs with respect to the user these values are RVs. We define the probability that a UAV $i$ is an interferer as 

\subsection{Conditional Coverage Probability}
%In this subsection we describe how the coverage probability for the UAV network is obtained as a function of the network design and wireless channel parameters.
\noindent
Deriving an expression for the coverage probability involves the intermediate steps of deriving an expression for the conditional coverage probability in terms of the Laplace transforms of the LOS and NLOS interferers, followed by deriving analytical expressions for these Laplace transforms. The conditional coverage probability is defined as the probability that the SINR of the downlink signal from the serving UAV to the user is above a threshold $\theta$, given $R_1=r_1$. Considering Nakagami-$m$ fading, the conditional coverage probability $\Pd(\sinr\geq \theta |R_1=r_1)$ is obtained following (21) in \cite{Chetlur_2017} as

\begin{equation}
 %\Pd(\sinr\geq \theta |R_1=r_1) =  \\
 \sum\limits_{n=0}^{m_{t_1}-1}\frac{s_{t_1}^n}{n!} (-1)^n \frac{d^n \Lc_{I}((p\eta(\omega))^{-1}s_{t_1})}{ds_{t_1}^n},
\end{equation}

\noindent
where $s_{t_1}= m_{t_1}\theta(r_1^2+\gamma^2)^{\alpha_{t_1}/2}$, $m_{t_1}$ is the Nakagami-$m$ fading term and $\Lc_{I}$ denotes the Laplace transform of the total interference. The LOS and NLOS interferers are distributed independently of one another, the proof of this is similar to the proof in \cite{Haenggi_2013} and is omitted here. Due to this, the Laplace transform above can be separated into a product of the Laplace transforms of the aggregate LOS and aggregate NLOS interference, along with the introduction of the noise-related term. This allows us to express the conditional coverage probability for an LOS serving UAV $\Pd(\sinr\geq \theta |R_1=r_1,t_1 = \text{L})$ as 

\begin{align}
&\sum\limits_{n=0}^{m_L-1}\frac{s_L^n}{n!} (-1)^n 
 \cdot \sum_{i_L+i_N+i_{\sigma}=n}\frac{n!}{i_L!i_N!i_{\sigma}!} \nonumber \\
 &\cdot(-(p\eta(\omega))^{-1}\sigma^2)^{i_{\sigma}}\exp(-(p\eta(\omega))^{-1}s_L\sigma^2) \nonumber \\
 &\cdot\frac{d^{i_L} \Lc_{I_L}((p\eta(\omega))^{-1}s_L)}{ds_L^{i_L}} \frac{d^{i_N}\Lc_{I_N}((p\eta(\omega))^{-1}s_L)}{ds_L^{i_N}},
\label{eq:condProb3}
\end{align}

\noindent
where $\Lc_{I_L}$ and $\Lc_{I_N}$ are the Laplace transforms of the aggregate LOS and NLOS interference, respectively, and the second sum is over all the combinations of non-negative integers $i_L,i_N$ and $i_{\sigma}$ that add up to $n$. The conditional coverage probability given an NLOS serving UAV $\Pd(\sinr\geq \theta |R_1=r_1,t_1 = \text{N})$ is calculated as in \Eq{condProb3} with $m_N$, $\alpha_N$ and $s_N$ replacing $m_L$, $\alpha_L$ and $s_L$.

\subsection{Laplace Transform of Aggregate Interference}
 The Laplace transform of the aggregate LOS interference $\Lc_{I_L}((p\eta(\omega))^{-1}s_L)$ given an LOS serving UAV is expressed as

\begin{align}
&\Ed_{\Phi^\prime_{\mathcal{W}L}}\bigg[\prod_{r\in\Phi^\prime_{\mathcal{W}L} \setminus [0,b_L(\text{L})]}\Ed_{H_L} \left[\exp\Big(-s_L H_L (r^2+\gamma^2)^{-\alpha_L/2}\Big)\right]\bigg]  \nonumber \\
&\overset{(a)}{=}\Ed_{\Phi^\prime_{\mathcal{W}L}}\left[\prod_{r\in\Phi^\prime_{\mathcal{W}L} \setminus [0,b_L(\text{L})]}g(r,s_L,m_L,\alpha_L)\right]  \nonumber \\
&\overset{(b)}{=}\exp\Bigg(-\int\limits_{b_L(\text{L})}^{\Rcone} \left(1-g(r,s_L,m_L,\alpha_L)\right) \lambda'_{L}(r) \dr r\Bigg) 
\label{eq:laplace}
\end{align}

\noindent
where 
\begin{equation}
 g(r,s_L,m_L,\alpha_L) = \left(\frac{m_L}{s_L(r^2+\gamma^2)^{-\alpha_L/2} + m_L}\right)^{m_L}   \nonumber,
\end{equation}

\noindent
 $(a)$ comes from Nakagami-$m$ fading having a gamma distribution, $(b)$ comes from the probability generating functional of the PPP \cite{Haenggi_2013} and $\lambda'_{L}(r) = \Pd_{LOS}(r) 2\pi\lambda r$. From the definition of the LOS probability function we can observe that $\Pd_{LOS}(r)$ is a step function. We use this fact to separate the integral above into a sum of weighted integrals, resulting in the following expression
 
 \begin{equation}
2\pi\lambda\sum\limits_{j=\floor*{b_L(\text{L})\sqrt{\beta\delta}}}^{\floor*{\Rcone\sqrt{\beta\delta}}} \Pd_{LOS}(l)\int\limits_{l}^{u}(1-g(r,s_L,m_L,\alpha_L))r\dr r 
\label{eq:laplaceSum}
 \end{equation}

\noindent
where $l = \max(b_L(\text{L}),j/\sqrt{\beta\delta})$ and $u = \min(\Rcone,(j+1)/\sqrt{\beta\delta})$.  The integral $\int\limits_{\Lower}^{\Upper}(1-g(r,s_L,m_L,\alpha_L))r\dr r$ can then be expressed as 

\begin{align}
%&\int\limits_{\Lower}^{\Upper}\left(1-\left(\frac{m_L}{s_L(r^2+\gamma^2)^{-\alpha_L/2} + m_L}\right)^{m_L}\right) r \dr r \nonumber \\
&\overset{(a)}{=} \int\limits_{(\Lower^2+\gamma^2)^{1/2}}^{(\Upper^2+\gamma^2)^{1/2}}\left(1-\left(\frac{m_L}{s_Ly^{-\alpha_L} + m_L}\right)^{m_L}\right)y\dr y \nonumber \\
%& \int\limits_{(\Lower^2+\gamma^2)^{1/2}}^{(\Upper^2+\gamma^2)^{1/2}}\left(1-\left(1-\frac{sy^{-\alpha_L}}{sy^{-\alpha_L} + 1}\right)^{m}\right)y\dr y \nonumber \\
%&= \int\limits_{(\Lower^2+\gamma^2)^{1/2}}^{(\Upper^2+\gamma^2)^{1/2}}\left(1-\left(1-\frac{1}{1 + y^{\alpha_L}m_Ls^{-1}}\right)^{m_L}\right)y\dr y \nonumber \\
%\end{align}
%\begin{align}
& \overset{(b)}{=} \frac{ 1}{\alpha_L}\int\limits_{(\Lower^2+\gamma^2)^{\alpha_L/2}}^{(\Upper^2+\gamma^2)^{\alpha_L/2}}\left(1-\left(1-\frac{1}{1 + zm_Ls^{-1}}\right)^{m_L}\right)z^{2/\alpha_L - 1}\dr z \nonumber \\
& \overset{(c)}{=} \frac{    1}{\alpha_L}\sum\limits_{k=1}^{m_L}\binom{m_L}{k}(-1)^{k+1}\int\limits_{(\Lower^2+\gamma^2)^{\alpha_L/2}}^{(\Upper^2+\gamma^2)^{\alpha_L/2}}\frac{z^{2/\alpha_L - 1}}{(1+zm_Ls_L^{-1})^k}\dr z ,\nonumber \\
%\label{eq:laplace_part2}
&\overset{(d)}{=} \frac{1}{2}\sum\limits_{k=1}^{m_L}\binom{m_L}{k}(-1)^{k+1}\nonumber \\
&\bigg((\Upper^2+\gamma^2)\mbox{$_2$F$_1$}\Big(k,\frac{2}{\alpha_L};1+\frac{2}{\alpha_L};  -\frac{m_L(\Upper^2+\gamma^2)^{\alpha_L/2}}{s_L}\Big) \nonumber\\
&-(\Lower^2+\gamma^2)\mbox{$_2$F$_1$}\Big(k,\frac{2}{\alpha_L};1+\frac{2}{\alpha_L};-\frac{m_L(\Lower^2+\gamma^2)^{\alpha_L/2}}{s_L}\Big)\bigg),
\label{eq:laplace_final}
\end{align}

\noindent
where $(a)$ stems from the substitution $y=(r^2+\gamma^2)^{1/2}$, $(b)$ from the substitution $z = y^{a_L}$, $(c)$ from applying binomial expansion and $(d)$ from using \cite{Ryzhik_2007}[Eq. 3.194.1], 
where $\mbox{$_2$F$_1$}(a,b;c;z)$ denotes the Gauss hypergeometric function. Inserting this solution into \Eq{laplaceSum} we obtain an expression for the Laplace transform of the LOS interferers \Eq{laplace}. Note that the Laplace transform for the NLOS interferers $\Lc_{I_N}((p\eta(\omega))^{-1}s_L)$ is solved by simply substituting $\lambda'_{L}(r)$ with $\lambda'_{N}(r)$, $b_L(\text{L})$ with $b_N(\text{L})$ and $g(r,s_L,m_L,\alpha_L)$ with $g(r,s_L,m_N,\alpha_N)$ in \Eq{laplace} and solving as shown. The above integration is for the case when the serving UAV is LOS; if the serving UAV is NLOS we substitute $s_L$ with $s_N$ as defined in the previous subsection and $b_L(\text{L})$ with $b_L(\text{N})$. The higher derivatives of the Laplace transforms become cumbersome to solve manually for larger values of the serving UAV fading parameter, so in order to obtain an analytical expression we treat the Laplace transforms as composite functions and apply Fa\`{a} di Bruno's formula for higher derivatives. 

\subsection{Coverage Probability}

%\begin{proposition}
To obtain the overall coverage probability for the reference user in the network we decondition the conditional coverage probability as defined in the previous section with respect to the indicator variable $t_1$ by multiplying by the two probability distributions given in \Eq{pdfLOS} and \Eq{pdfNLOS}, we then decondition with respect to the horizontal distance random variable $R_1$ via integration. $\Pd(\sinr\geq \theta)$ is given as
%\end{proposition}

\begin{multline}
\int\limits_{0}^{\Rcone}\bigg(\Pd(\sinr\geq \theta |R_1=r_1,t_1 = \text{L})f_{R_{1},t_1}(r_1,t_1 = \text{L}) \\
+\Pd(\sinr\geq \theta |R_1=r_1,t_1 = \text{N})f_{R_{1},t_1}(r_1,t_1 = \text{N})\bigg) \dr r_1
 \label{eq:pcov_final}
\end{multline}

\section{Numerical Results}

\begin{table}[b!]
\begin{center}
\caption{Numerical Result Parameters}
\begin{tabular}{ |c|c| } 
 \hline
 Parameter & Value  \\ 
 \hline
 $\omega$ & \unit[$2.87$]{rad} \\
 $\alpha_L$ & 2.1 \\
 $\alpha_N$ & 4 \\
 $m_L$ & 3 \\
 $m_N$ & 1 \\
 $p$ & \unit[0.1]{W} \\
 $\sigma^2$ & \unit[$10^{-9}$]{W} \\
 $\beta$ & \unit[300]{$/\text{km}^2$}\\
 $\delta$ & 0.5\\
 $\kappa$ & \unit[50]{m} \\
 \hline
\end{tabular}
 \label{tab:table}
\end{center}
\end{table}

\begin{figure}[t!]
\centering
	\includegraphics[width=.40\textwidth]{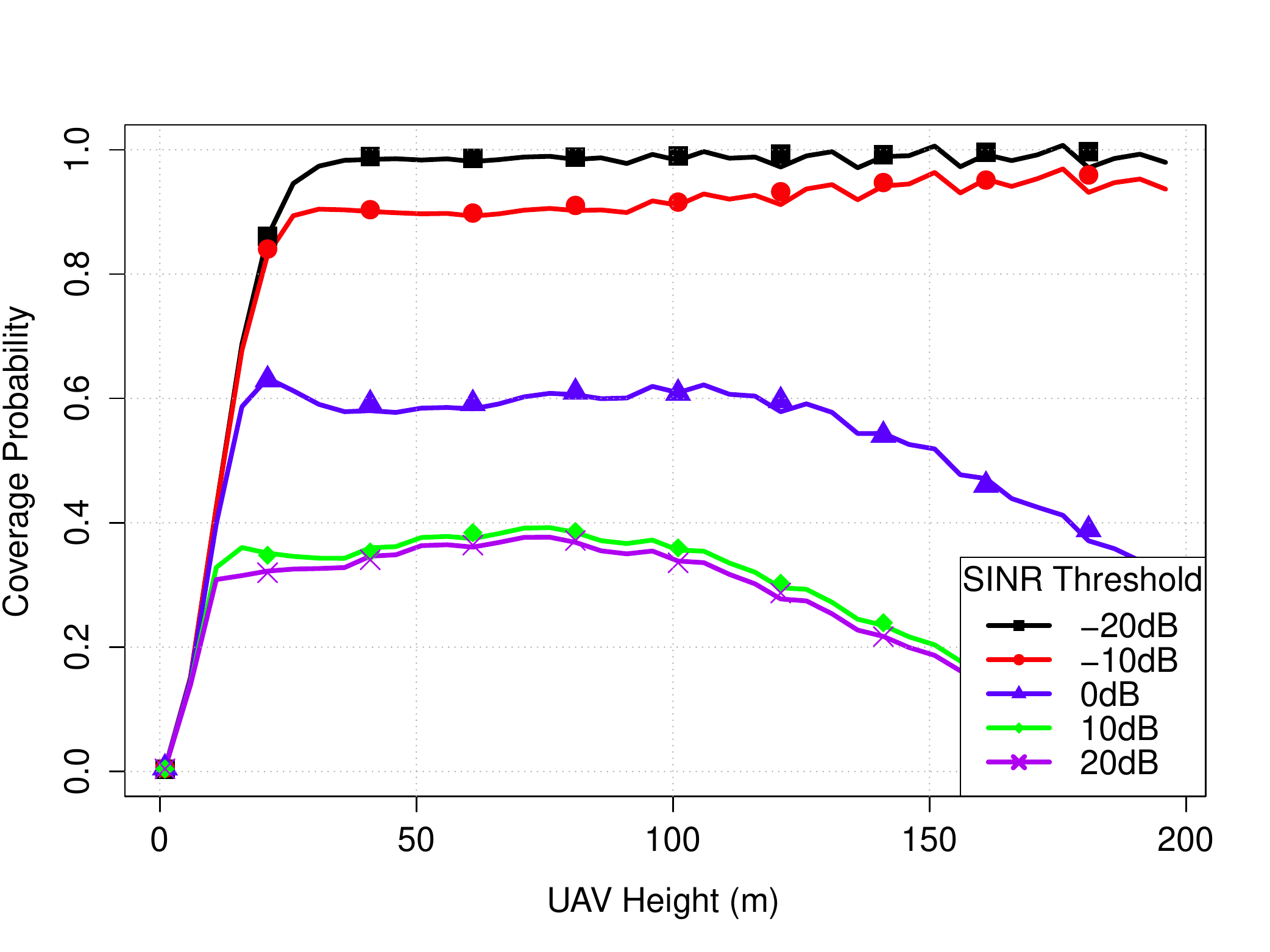}
	\caption{
	Coverage probability given a UAV density of \unit[25]{$\text{/km}^2$}
	\vspace{-5mm}
	}
	\label{fig:LOSSINR}
\end{figure}

\begin{figure}[t!]
\centering
	\includegraphics[width=.40\textwidth]{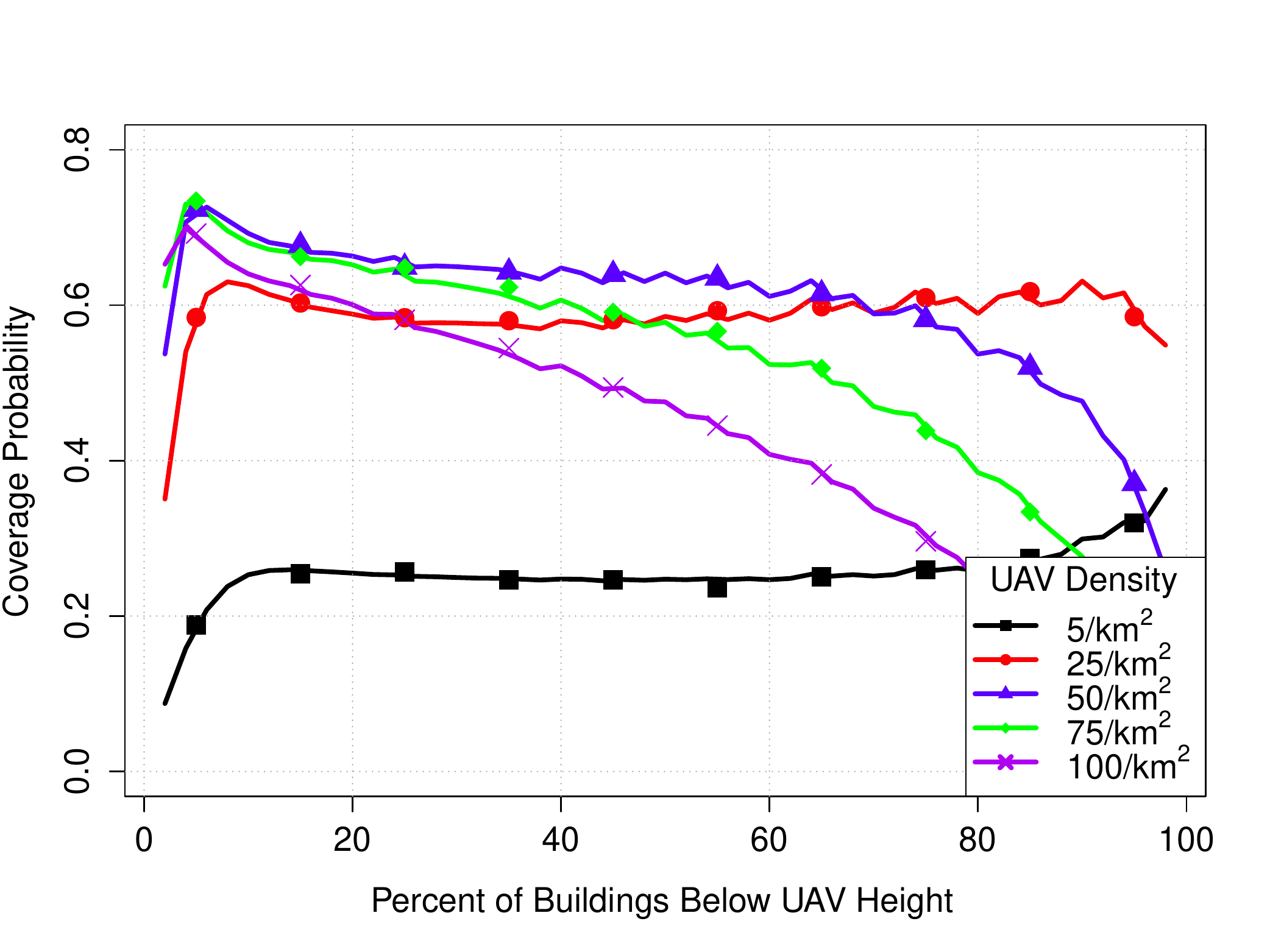}
	\caption{
	Coverage probability for multiple densities given a threshold of \unit[0]{dB}
	\vspace{-5mm}
	}
	\label{fig:LOSDensity}
\end{figure}

\begin{figure}[t!]
\centering
	\includegraphics[width=.40\textwidth]{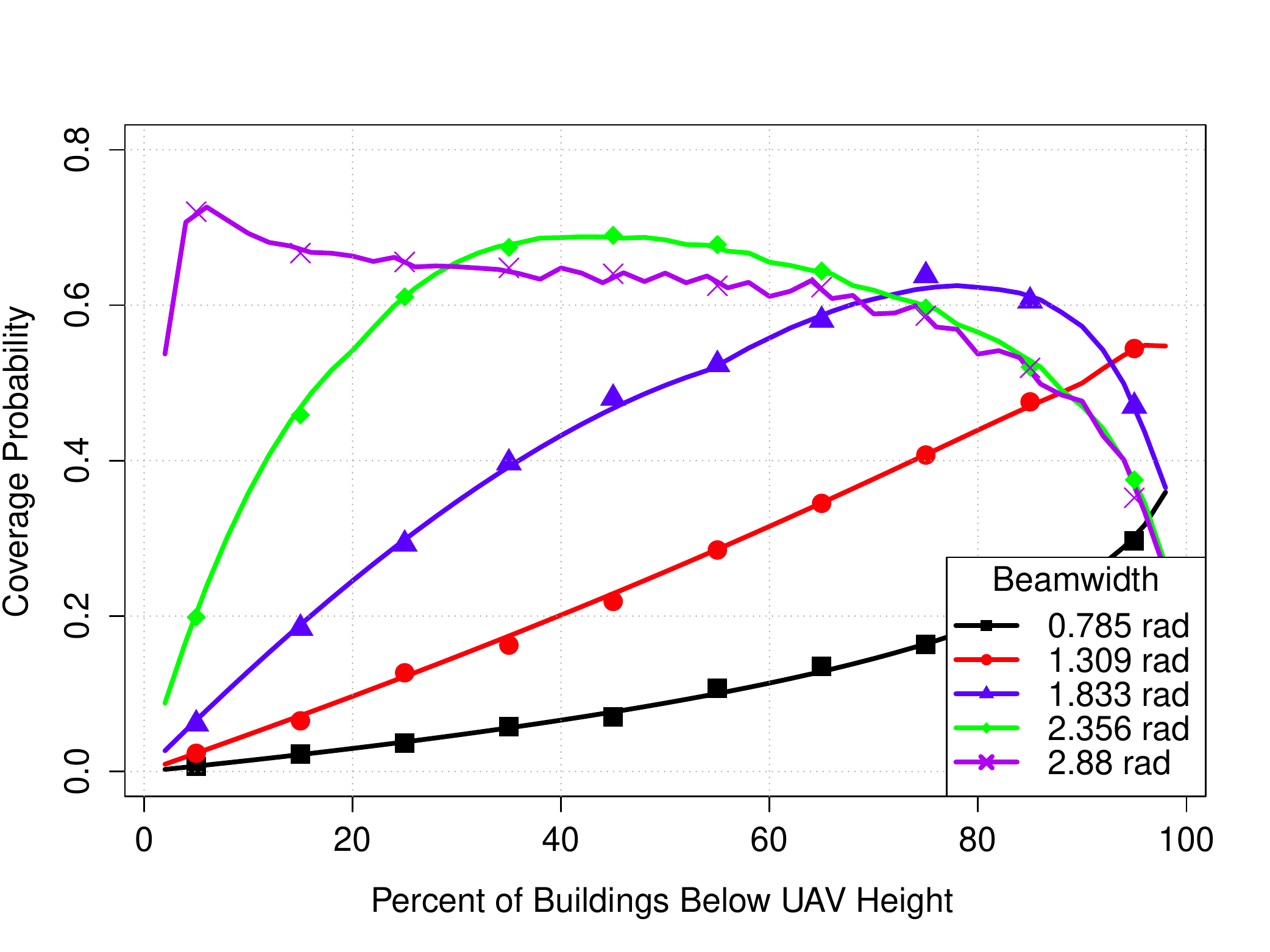}
	\caption{
	Coverage probability given a threshold of \unit[0]{dB} and UAV density of \unit[50]{$\text{/km}^2$}
	\vspace{-5mm}
	}
	\label{fig:LOSBeamwidth}
\end{figure}

\begin{figure}[t!]
\centering
	\includegraphics[width=.40\textwidth]{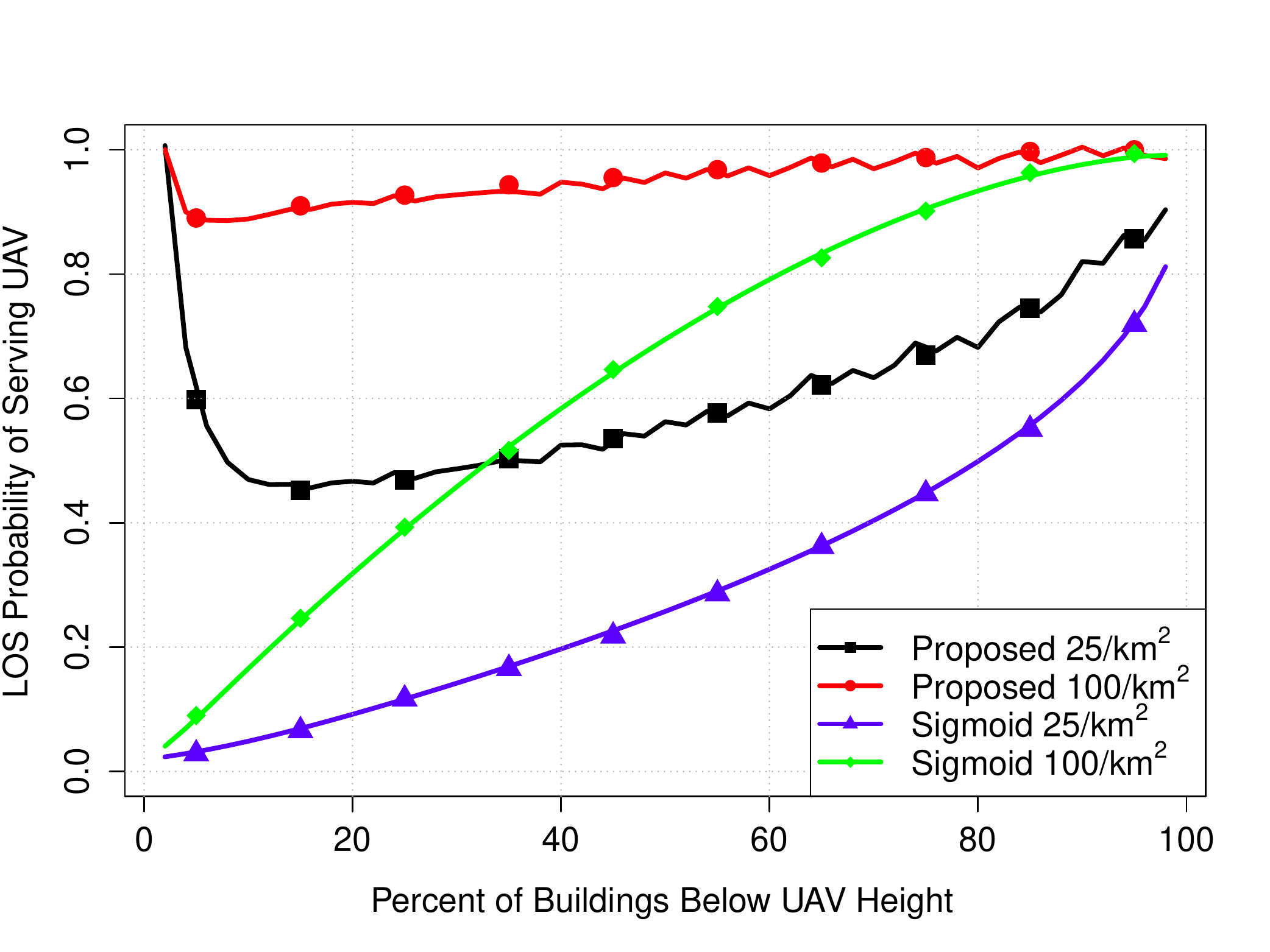}
	\caption{
	Probability of an LOS channel between a user and its serving UAV
	\vspace{-5mm}
	}
	\label{fig:Comparison}
\end{figure}

In this section we demonstrate how our model can provide insight into the behaviour of low-altitude UAV networks in urban environments. In Figures \ref{fig:LOSSINR} to \ref{fig:Comparison}, solid lines denote the analytical values for the coverage probability (from \Eq{pcov_final}) and the markers denote results from Monte Carlo trials. Unless stated otherwise the parameters used for the numerical results are from Table \ref{tab:table}. %The transmit power of the UAVs and the noise power are taken as \unit[0.1]{W} and \unit[$10^{-9}$]{W}, respectively, with the wireless channel parameters $\alpha$ and $m$ being set to 2.1 and 3, respectively. In \Cref{fig:assProb,fig:NakSINR,fig:NakDensityN,fig:NakAngleN}, solid lines denote the analytical values for the coverage probability (from \Eq{pcov_final}) and the markers denote results from Monte Carlo trials.

\Fig{LOSSINR} shows the effect of varying the UAV height on the coverage probability, given different SINR threshold values. We can see that initially the coverage probabilities for all the SINR thresholds improve as we increase the height. This is due to the UAVs increasing their coverage areas, which maximises the probability that there is at least one UAV within range of the user and providing sufficient SINR. Past a certain height; however, the curves show different behaviour: for a high SINR threshold, coverage probability monotonically decreases with increased height. This shows how the signals are more vulnerable to the increasing number of LOS interferers that appear as the UAV heights increase. The coverage curves considering a low SINR threshold, however, show different behaviour: the coverage probability appears to monotonically increase. This suggests that while the low SINR signals are negatively affected by the increasing number of LOS interferers, the increasing probability that the serving UAV will be within LOS causes an overall increase in the coverage probability. Additionally, note how the curves appear to have minor ripples: this is due to the complex building grid environment which causes micro-fluctuations in the performance as we vary the UAV locations. %This captures the sort of performance fluctuations we may expect to see in real-world, non-idealised wireless environments.

In \Fig{LOSDensity} we consider how increasing the UAV height to be above a certain percentage of the buildings affects the coverage probability, for UAV networks of different densities. The figure shows two very different behaviours: the coverage probability for the lowest density network appears to modestly improve as we increase the UAV height, whereas an increase in height deteriorates the coverage of the high density networks. This is explained by considering the effect of the buildings on the networks. At low densities the serving UAV for the reference user may be concealed behind several buildings, and increasing the UAV height increases the chances of establishing an LOS channel. The low number of interferers within range means that as the channel between a user and its serving UAV improves, the net impact on the network performance is positive. Note that the overall coverage probability for the low density network is low precisely because so few UAVs cover the operating area, resulting in a low probability of a UAV being within range of the user. In a high density network the serving UAV to a user is likely to be close enough that there are few buildings to interfere with the signal. The buildings in this scenario do not impede the serving UAV signal but instead shield the user from interfering UAVs a further distance away. Increasing the UAV height then will expose the user to these interferers while at the same time worsening the serving signal, resulting in a drop in coverage.

In \Fig{LOSBeamwidth} we demonstrate the effect of the UAV antenna beamwidth on the coverage probability. The  coverage curves suggest that narrower beamwidths perform best at larger UAV heights. This is due to the effect of the beawmidths on the probability of the user being within range of a UAV: narrow beamwidth UAVs create a narrow coverage cone and as a result must operate at larger heights to ensure that users can be within range of service. We see that each beamwidth value has an associated optimum UAV height for a given SINR threshold, UAV density and building environment.

In \Fig{Comparison} we show the probability that a user that is within range of the network will have an LOS link to the serving UAV under our LOS model and the sigmoid approximation adopted in \cite{AlHourani_2014}, given two different UAV densities. The sigmoid model gives the LOS probability as a function of the vertical angle between the UAV and the user, and as such when the UAV is close to the ground the LOS probability to its users approaches zero, despite the fact that the UAV coverage cone is very small and therefore the users are very close to the UAV. As the height increases this probability steadily improves due to the increasing vertical angle. Our model captures a more realistic behaviour of the LOS probability; when the UAV is very low to the ground, due to the size of its coverage cone its users are close enough that no LOS-blocking buildings are in the way, ensuring an LOS probability approaching unity. As the height increases the increasing coverage cone allows users further away to associate to the UAV, resulting in more users behind buildings being served by the UAV, which negatively affects the LOS probability. Finally, as the UAV ascends above the majority of buildings this LOS probability steadily improves to reflect the fact that there will be fewer buildings tall enough to block the UAV-user link.

%\begin{figure}[t!]
 %centering
%	\includegraphics[width=.45\textwidth]{NakParameterN.eps}
%	\caption{
%	Coverage probability given a threshold of \unit[-10]{dB}, UAV beamwidth of $\pi/2$, pathloss exponent of 2.1, and UAV density of \unit[20]{$\text{/km}^2$}. 
%	\vspace{-5mm}
%	}
%	\label{fig:NakParameterN}
%\end{figure}

\section{Conclusion}
In this paper we have used stochastic geometry to model a UAV network in an urban environment, considering UAV network parameters such as density and height above ground, as well as environment parameters such as the building density and building heights. We derived an expression for the coverage probability of the UAV network as a function of these parameters and then verified the derivation numerically, while showing the trade-offs in performance that occur under different network conditions.

\ifCLASSOPTIONcaptionsoff
  \newpage
\fi



\bibliographystyle{./bib/IEEEtran}
\bibliography{./bib/IEEEabrv,./bib/IEEEfull}

\begin{thebibliography}{10}
\providecommand{\url}[1]{#1}
\csname url@samestyle\endcsname
\providecommand{\newblock}{\relax}
\providecommand{\bibinfo}[2]{#2}
\providecommand{\BIBentrySTDinterwordspacing}{\spaceskip=0pt\relax}
\providecommand{\BIBentryALTinterwordstretchfactor}{4}
\providecommand{\BIBentryALTinterwordspacing}{\spaceskip=\fontdimen2\font plus
\BIBentryALTinterwordstretchfactor\fontdimen3\font minus
  \fontdimen4\font\relax}
\providecommand{\BIBforeignlanguage}[2]{{%
\expandafter\ifx\csname l@#1\endcsname\relax
\typeout{** WARNING: IEEEtran.bst: No hyphenation pattern has been}%
\typeout{** loaded for the language `#1'. Using the pattern for}%
\typeout{** the default language instead.}%
\else
\language=\csname l@#1\endcsname
\fi
#2}}
\providecommand{\BIBdecl}{\relax}
\BIBdecl

\bibitem{Motlagh_2016}
N.~H. Motlagh, T.~Taleb, and O.~Arouk, ``{Low-Altitude Unmanned Aerial
  Vehicles-Based Internet of Things Services: Comprehensive Survey and Future
  Perspectives},'' \emph{IEEE Internet of Things Journal}, vol.~3, no.~6, pp.
  899--922, Dec 2016.

\bibitem{Atkins_2010}
E.~M. Atkins, ``{Risk Identification and Management for Safe UAS Operation},''
  in \emph{3rd International Symposium on Systems and Control in Aeronautics
  and Astronautics}, June 2010, pp. 774--779.

\bibitem{AlHourani_20142}
A.~Al-Hourani, S.~Kandeepan, and A.~Jamalipour, ``{Modeling Air-to-Ground Path
  Loss for Low Altitude Platforms in Urban Environments},'' in \emph{IEEE
  Globecom}, Dec 2014, pp. 2898--2904.

\bibitem{AlHourani_2014}
A.~Al-Hourani, S.~Kandeepan, and S.~Lardner, ``{Optimal LAP Altitude for
  Maximum Coverage},'' \emph{IEEE Wireless Communications Letters}, vol.~3,
  no.~6, pp. 569--572, Dec 2014.

\bibitem{Mozaffari_2015}
M.~Mozaffari \emph{et~al.}, ``{Drone Small Cells in the Clouds: Design,
  Deployment and Performance Analysis},'' in \emph{IEEE Globecom}, Dec 2015,
  pp. 1--6.

\bibitem{Hayajneh_2016}
A.~M. Hayajneh \emph{et~al.}, ``{Optimal Dimensioning and Performance Analysis
  of Drone-Based Wireless Communications},'' in \emph{IEEE Globecom Workshops},
  Dec 2016, pp. 1--6.

\bibitem{Mozaffari_20162}
M.~Mozaffari \emph{et~al.}, ``{Mobile Internet of Things: Can UAVs Provide an
  Energy-Efficient Mobile Architecture?}'' in \emph{IEEE Globecom}, Dec 2016,
  pp. 1--6.

\bibitem{Fotouhi_2016}
A.~Fotouhi, M.~Ding, and M.~Hassan, ``{Dynamic Base Station Repositioning to
  Improve Performance of Drone Small Cells},'' in \emph{IEEE Globecom
  Workshops}, Dec 2016, pp. 1--6.

\bibitem{Ravi_2016}
V.~V.~C. Ravi and H.~S. Dhillon, ``{Downlink Coverage Probability in a Finite
  Network of Unmanned Aerial Vehicle (UAV) Base Stations},'' in \emph{17th IEEE
  International Workshop on Signal Processing Advances in Wireless
  Communications (SPAWC)}, July 2016, pp. 1--5.

\bibitem{Chetlur_2017}
V.~{Vardhan Chetlur} and H.~S. {Dhillon}, ``{Downlink Coverage Analysis for a
  Finite 3D Wireless Network of Unmanned Aerial Vehicles},'' \emph{ArXiv
  e-prints}, Jan. 2017.

\bibitem{Zhang_2017}
C.~Zhang and W.~Zhang, ``{Spectrum Sharing for Drone Networks},'' \emph{IEEE
  Journal on Selected Areas in Communications}, vol.~35, no.~1, pp. 136--144,
  Jan 2017.

\bibitem{Balanis_2005}
C.~A. Balanis, \emph{Antenna Theory: Analysis and Design}.\hskip 1em plus 0.5em
  minus 0.4em\relax Wiley-Interscience, 2005.

\bibitem{ITUR_2012}
``{Recommendation P.1410-5 "Propagation Data and Prediction Methods Required
  for the Design of Terrestrial Broadband Radio Access Systems Operating in a
  Frequency Range From 3 to 60 GHz"},'' ITU-R, Tech. Rep., 2012.

\bibitem{Haenggi_2013}
M.~Haenggi, \emph{{Stochastic Geometry for Wireless Networks}}.\hskip 1em plus
  0.5em minus 0.4em\relax Cambridge University Press, 2013.

\bibitem{Ryzhik_2007}
I.~Gradshteyn and I.~Ryzhik, \emph{Table of Integrals, Series, and
  Products}.\hskip 1em plus 0.5em minus 0.4em\relax Academic Press, 2007.

\end{thebibliography}
\end{document}